\documentclass[%
reprint,
amsmath,amssymb,
aps,
prl,
]{revtex4-1}

\usepackage{graphicx}% Include figure files
\usepackage{dcolumn}% Align table columns on decimal point
\usepackage{bm}% bold math
\usepackage{amsmath}
\usepackage{subfigure}
\usepackage{epstopdf}
\usepackage{csvsimple,longtable,booktabs}
\usepackage{array}
\usepackage{tabularx}
\newcolumntype{C}[1]{>{\centering\let\newline\\\arraybackslash\hspace{0pt}}m{#1}}

\usepackage{hyperref}% add hypertext capabilities
\usepackage[mathlines]{lineno}% Enable numbering of text and display math
\usepackage{color,soul}

\usepackage{enumerate}
\usepackage{enumitem}
\usepackage{newlfont}
\usepackage{tikz-network}

\usepackage{tikz}

\def\genbox#1#2#3#4#5#6{% #1=0/1, #2=color, #3=shape, #4=raise, #5=width, #6=width/2
	\leavevmode\raise#4bp\hbox to#5bp{\vrule height#5bp depth0bp width0bp
		\pdfliteral{q .5 w \csname #2COLOR\endcsname\space RG
			\csname #3PDF\endcsname{#5}{#6} S Q
			\ifx1#1 q \csname #2COLOR\endcsname\space rg 
			\csname #3PDF\endcsname{#5}{#6} f Q\fi}\hss}}

\begin{document}
	
	\preprint{APS/123-QED}
	
	\title{Cohesion and segregation in higher-order networks}
	
	\author{Demival Vasques Filho}
	%\altaffiliation[]{Physics Department, XYZ University.}%Lines break automatically or can be forced with \\
	\email{vasquesfilho@ieg-mainz.de}
	\affiliation{%
		%Digital Historical Research, 
		Leibniz-Institut f\"ur Europ\"aische Geschichte\\
		Alte Universit\"atsstra{\ss}e 19, 55116 Mainz, Germany
	}%
	
	\date{\today}% It is always \today, today,
	
\begin{abstract}
Looking to overcome the limitations of traditional networks, the network science community has lately given much attention to the so-called higher-order networks, where \textit{group} interactions are modeled alongside \textit{pairwise} ones. %The three most common ways of representing such interactions are through bipartite (two-mode) networks, hypergraphs, and simplicial complexes.
While degree distribution and clustering are the most important features of traditional network structure, higher-order networks present two additional fundamental properties that are barely addressed: the group size distribution and overlaps. Here, I investigate the impact of these properties on the network structure, focusing on cohesion and segregation (fragmentation and community formation). For that, I create artificial higher-order networks with a version of the configuration model that assigns degree to nodes and size to groups and forms overlaps with a tuning parameter $p$. Counter-intuitively, the results show that a high frequency of overlaps favors both network cohesion and segregation---the network becomes more modular and can even break into several components, but with tightly-knit communities.

%\begin{description}
%\end{description}
\end{abstract}
	
%\pacs{Valid PACS appear here}% PACS, the Physics and Astronomy Classification Scheme.
%\keywords{Suggested keywords}%Use showkeys class option if keyword display desired
\maketitle

%\section{\label{sec:introduction}Introduction}

%cohesion (local clustering), defined as a tightly knit set of actors around the ego, and fragmentation, defined as the $F$ measure.

Systems of different nature present not only \textit{pairwise} interactions involving only two elements of the system but also \textit{group} interactions that involve three or more elements. Traditionally, the latter have been represented as a collection of the former, even though network techniques to explicitly represent group interactions---namely bipartite graphs and hypergraphs---have been around for decades. Nevertheless, the relevance of modeling group interactions in the structure and dynamics of complex systems has only recently become pronounced; the study of these networks picked up pace, and the term \textit{higher-order networks}  was coined. Higher-order networks are, thus, a generalization of traditional networks with which we can model interactions of groups of any size (including size two)~\cite{vasques2020transitivity}.  

Using either bipartite graphs or hypergraphs depends on the system and the problem to be addressed. We can quickly transform one into the other but must keep in mind that they can provide some different topological information~\cite{estrada2006subgraph}. The main distinctions between them, however, lie in the mathematical representations and analytical solutions, while the concepts are mostly translatable.  

A bipartite network is a graph $B = \{U,V,L\}$, where $U$ and $V$ are disjoint sets of nodes and $L = \{(u,v):u \in U, v \in V\}$ is the set of edges connecting nodes of the different sets. We refer to the sets $U$ and $V$ as the bottom and top node sets, respectively. A hypergraph is a graph $H = \{U,E\}$, where $U$ is the set of nodes and $E = \{e_{1}...e_{m}\}$ is the set of hyperedges connecting non-empty subsets of $U$. The transformation is simple: the node set $U$ can be seen as the same for both cases; $E = V$, such that each hyperedge $e \in E$ corresponds to a top node $v \in V$; and hyperedge $e = v$ contains node $u$ if there exist an edge $(u,v) \in L$. From now on, I will use the hypergraph notation. 

The main structural properties of higher-order networks are:
\begin{itemize}[noitemsep]
	\item the degree distribution $P_{u}(k)$, where $k$ is the node degree or the number of hyperedges adjacent to the node,
	\item the group size distribution $P_{e}(d)$, where $d$ is the hyperedge size, or the number of nodes the hyperedge connects, and
	\item overlaps, when two nodes are connected by more than one hyperedge~\cite{vasques2018degree,vasquesfilho2019structure}.
\end{itemize}

These three properties determine the neighborhood size $q_{u}$ of node $u$ according to~\cite{vasques2018degree,vasques2020transitivity}:
\begin{equation}
	\label{eq:qi}
	q_{u} \leq \sum_{j=1}^{k_{u}}(d_{v_j} - 1)\,.
\end{equation}
That is, the neighborhood size of a node is equal to or less than the sum of the size (minus the node itself) of all hyperedges adjacent to that node. The equality (the upper bound) happens where there is no overlap in the network that contains node $u$. The higher the number of overlaps containing $u$ is, the smaller $u$'s expected neighborhood is. 

The idea of overlaps in networks representing higher-order interactions is not new, especially concerning bipartite networks. They are known as bipartite clustering~\cite{lind2005cycles,zhang2008clustering}, butterflies~\cite{sanei2018butterfly,wang2019vertex}, and four-cycles~\cite{wang2013exponential,vasques2020role}. Moreover, they are highly frequent in empirical networks~\cite{koskinen2012modelling,vasques2020role,vasques2020transitivity}. However, despite the much-needed attention that higher-order networks have attracted recently, the effect of overlaps in the network structure is little explored. Neglecting overlaps in these networks is similar to neglecting triangles in traditional networks.

Thus, the goal of this study is to examine how these overlaps, together with $P_{u}(k)$ and $P_{e}(d)$, affect cohesion, and segregation (fragmentation and community formation) in higher-order networks. To do so, I use a version of the configuration model---assigning degree to nodes and size to hyperedges from specific probability distributions---with several combinations of distributions, ranging from peaked to heavy-tailed~\cite{vasques2018degree}. These combinations are also used to explore how the shape of the probability distributions affects the structure of networks with arbitrary degree and group size distributions.

However, the frequency of overlaps in configuration models is negligible when the density of these artificial networks is comparable to that of empirical ones~\cite{vasques2020role,vasques2020transitivity}. To overcome this issue, I adapt the configuration model to create overlaps according to a tuning parameter $p$~\cite{vasques2020mechanisms}.  

The random process starts after assigning degree to nodes and size to hyperedges, without connecting them. The model picks a node (at random) and, with a probability $p$, tries to form an overlap. It does so by checking whether there are hyperedges and nodes with ``capacity'' to form an overlap, based on the size and degree assigned to them, respectively. If the model finds hyperedges and nodes with such ``capacity'', it randomly chooses two of these hyperedges and one of these nodes and connects them to the initial node to form an overlap. Otherwise (i.e., the algorithm does not find nodes and hyperedges with such ``capacity''), the model connects the initial node to another (random) node through a (random) hyperedge. The model repeats this process until no further nodes are available. 

\begin{figure*}
	\centering
	\subfigure{\label{fig:p_delta} \includegraphics[scale=0.55]{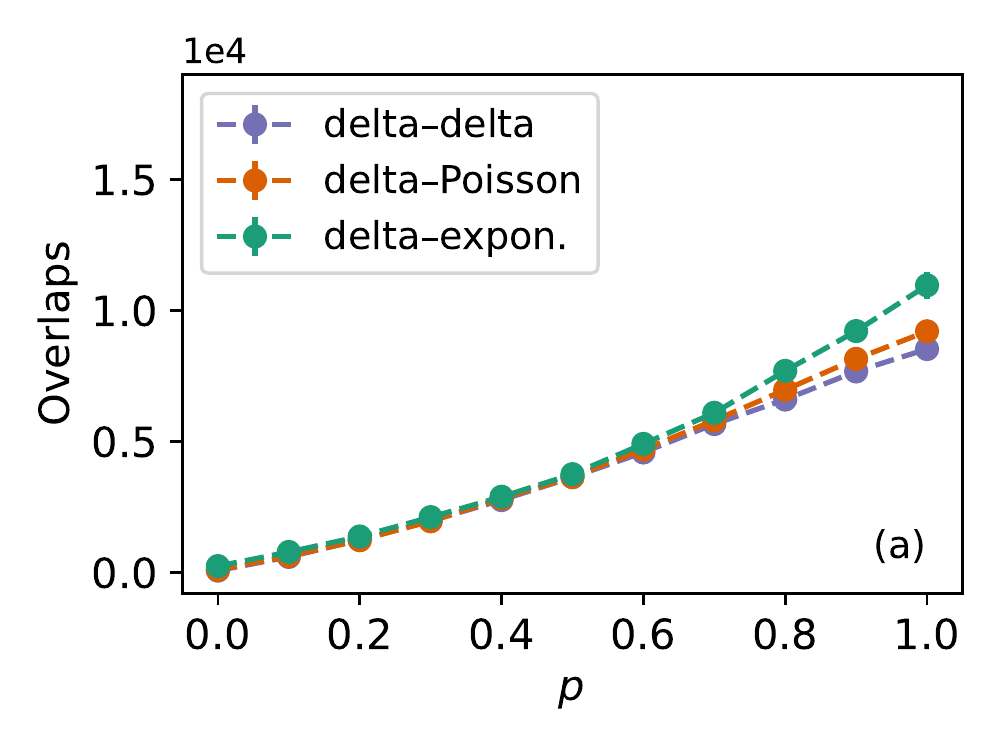}}
	\subfigure{\label{fig:p_poisson} \includegraphics[scale=0.55]{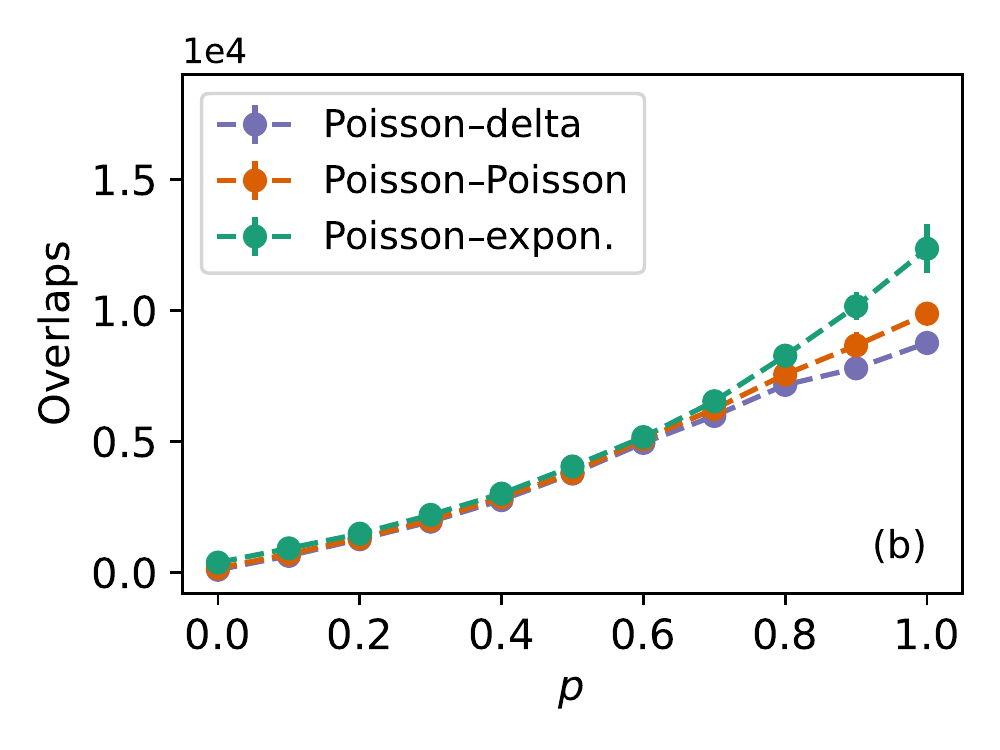}}
	\subfigure{\label{fig:p_exp} \includegraphics[scale=0.55]{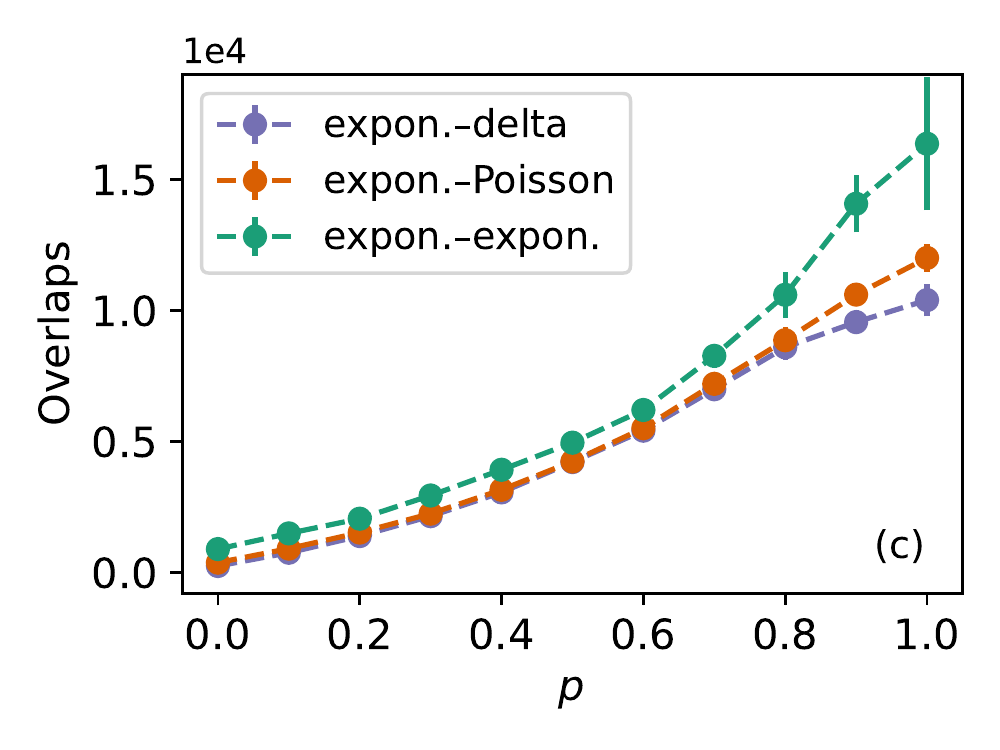}}
	\caption{Frequency of overlaps in higher-order networks with arbitrary distributions following the scheme hyperedge size---degree distributions, with (a) delta function, (b) Poisson, and (c) exponential distributions as hyperedge size distributions. Heavy-tailed distributions generate more overlaps due to the higher probability of two nodes being connected by the same hyperedge when they have a higher degree and a larger size, respectively, in random networks~\cite{vasques2020role}. The number of overlaps is the mean value over ten runs, and bars are the standard deviation.}
	\label{fig:overlap_vs_p}
\end{figure*}

The artificial networks are created following combinations of node degree and hyperedge size distributions using the delta function, Poisson, and exponential distributions, with $|U| \approx |E| \approx 1000$, and $\langle k \rangle \approx \langle d \rangle \approx 5$ (the values oscillate a little due to the model's stochasticity). Figure~\ref{fig:overlap_vs_p} shows the expected number of overlaps for the networks with each combination. Heavy-tailed distributions generate more overlaps for the same $p$ due to the higher probability of two nodes being connected by the same hyperedge (Eq. (15) of Ref.~\cite{vasques2020role}). 

Then, let us look at how overlaps affect the structure of higher-order networks. To measure the overall cohesion in the network, we will take the mean of the local clustering coefficient according to
\begin{equation}
	\label{eq:av_clust}
	\langle c \rangle = \frac{1}{|U|} \sum_{u \in U} cc_{u}\,,
\end{equation}
where $|U|$ is the number of nodes in the network. The local clustering coefficient $cc_{u}$ is given by
\begin{equation}
	\label{eq:loc_clust}
	cc_{u} = \dfrac{2E_{u}}{q_{u}(q_{u}-1)}\,,
\end{equation}
where $E_{u}=\{\{u',u''\} \subseteq N(u):~ (u',u'') \in E\}$ is the number of existing links between the $q_{u}$ nodes in the neighborhood $N(u)$ of $u$. That is, $cc_{u}$ is the density of $N(u)$. Note that this definition for clustering coefficient is that of traditional networks, introduced in Ref.~\cite{watts1998collective}, and links refer to pairwise relations between nodes. Although many works exist addressing clustering coefficients in hypergraphs~\cite{estrada2006subgraph,latapy2008basic,blond2005clustering,klamt2009hypergraphs}, I argue that we still should rely on the pairwise count of connections between nodes for addressing network cohesion. Thus, we calculate Eq.~\ref{eq:loc_clust} with a projection of the hypergraph (similarly to the well-known projection of bipartite networks). Counting $E_{u}$ is equivalent to counting the number of triangles in the neighborhood of $u$, centered in $u$, in the projection. 

Alternatively, in a hypergraph, we can walk closed paths of size three (i.e., triangles) following two ways. First, through three different hyperedges, forming the so-called hypertriangles, equivalent to six-cycles in bipartite networks~\cite{opsahl2013triadic,vasques2020role,vasques2020transitivity}. Second, within the same hyperedge of size equal to or larger than three, forming ``false'' hypertriangles~\cite{estrada2006subgraph}. Then, we achieve the same result as in Eq.~\ref{eq:loc_clust} by counting hypertriangles and ``false'' hypertriangles centered in $u$ in the hypergraph, excluding when the former are embedded in the latter~\cite{vasques2020transitivity}. The previous method of calculating clustering as a proxy of cohesion is simpler and more intuitive.

\begin{figure*}
	\centering
	\subfigure{\label{fig:cc_exp} \includegraphics[scale=0.55]{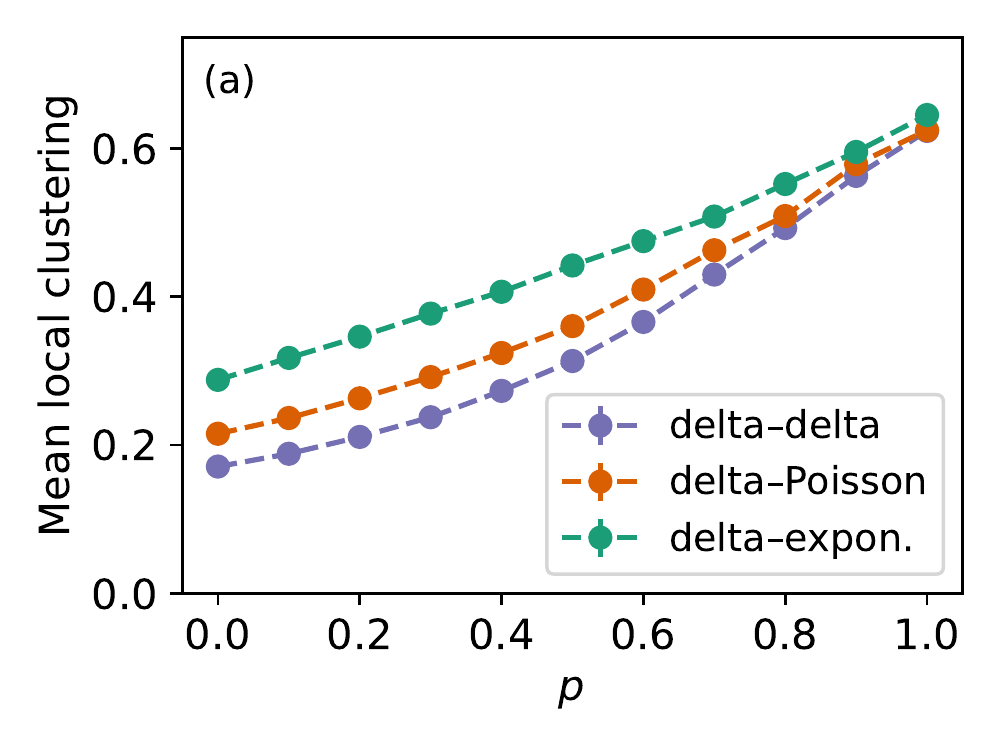}}
	\subfigure{\label{fig:cc_exp} \includegraphics[scale=0.55]{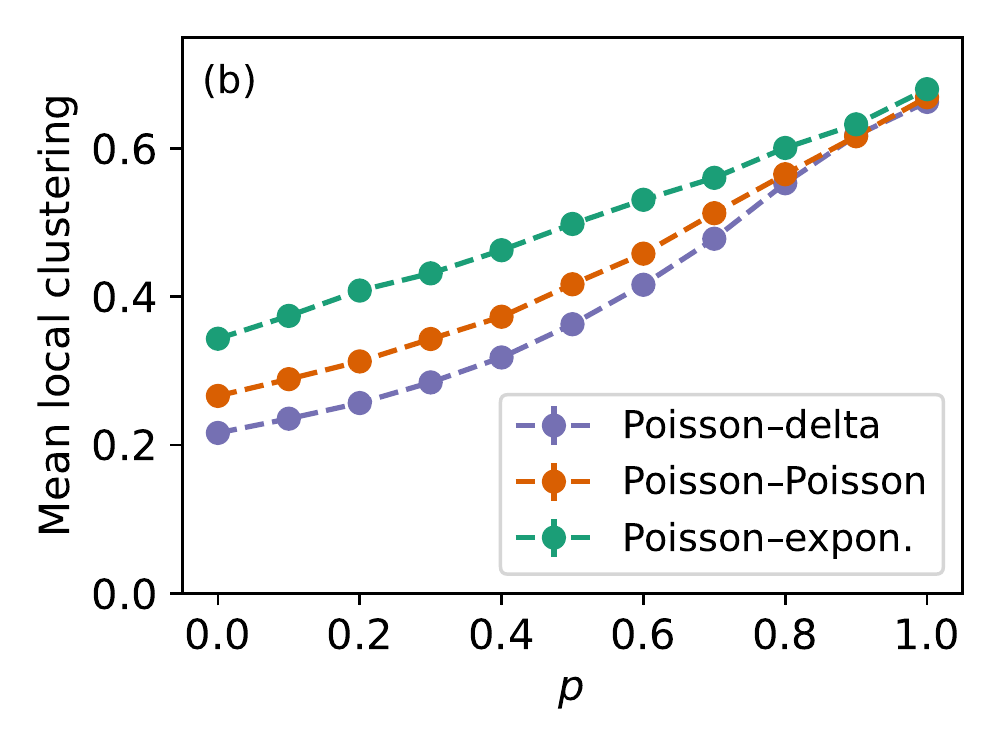}}
	\subfigure{\label{fig:cc_exp} \includegraphics[scale=0.55]{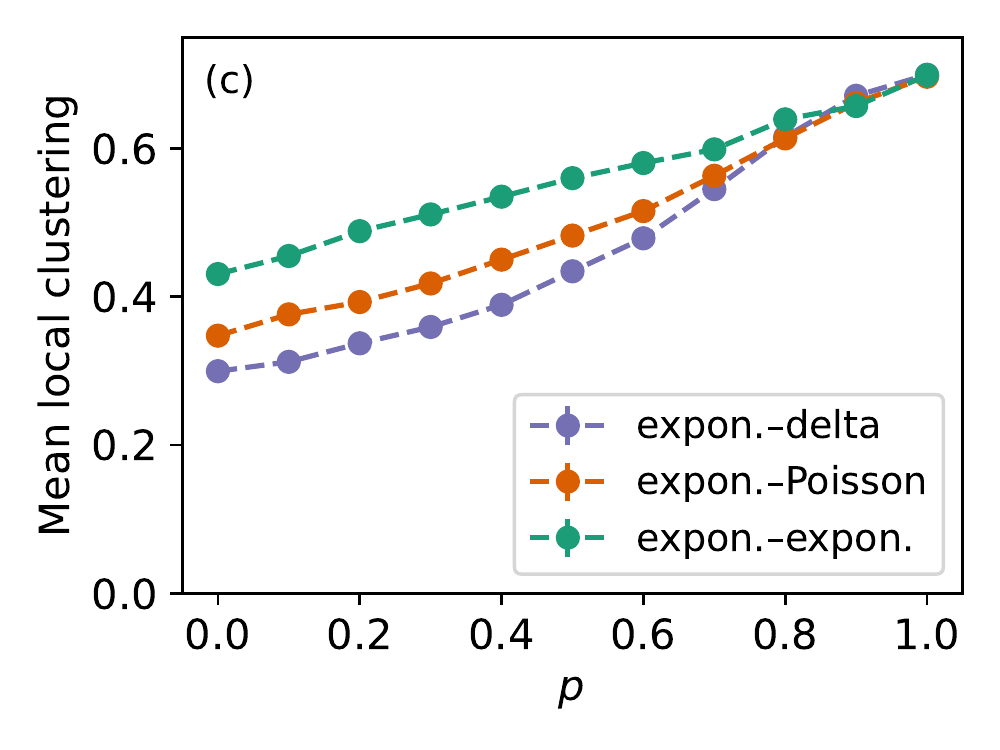}}
	\caption{Mean local clustering of the whole network as a function of $p$ for all $P_{e}(d)$-$P_{u}(k)$ combinations with (a) delta function, (b) Poisson, and (c) exponential hyperedge size distributions. The number of overlaps significantly increases cohesion (measured as local clustering). With high overlapping levels, cohesion becomes less dependent of node degree distributions. Values are the mean over ten runs. Standard deviation bars are unnoticeable.}
	\label{fig:clustering_vs_p}
\end{figure*}

An increasing number of overlaps favors cohesion in the whole network (Fig.~\ref{fig:clustering_vs_p}). The reason for that is actually straightforward: more overlaps result in smaller neighborhood sizes $q_{u}$ (Eq.~(\ref{eq:qi})) and, consequently, lower values  in Eq.~\ref{eq:loc_clust}.

Moreover, note how broader node degree and hyperedge size distributions also favor cohesion in the configuration model, albeit only slightly (compared to the effects of the parameter $p$). The longer distribution tails increase the probability of pairs of nodes connected through more than one hyperedge, generating some overlaps themselves~\cite{vasques2020role}, as Fig.~\ref{fig:p_exp} shows. That also explains the less pronounced effects of  the distribution tail for high values of $p$, especially concerning degree distributions. The configuration models reach a ``saturation'' level (with so many overlaps) such that nodes with high degrees (found in heavy-tailed distributions) have neighborhoods of similar size as that of nodes with degrees near the average (as the majority in peaked distributions).   

Next, let us look at segregation starting with network fragmentation. For that, we take Borgatti's $F$ measure~\cite{borgatti2006identifying} given by
\begin{equation}
	F = 1 - \frac{\sum_{i} s_{i}(s_{i}-1)}{|U|(|U|-1)}\,,
\end{equation} 
where $s_{i}$ is the size of component $i$ and $|U|$, again, is the number of nodes in the network.

\begin{figure*}
	\centering
	\subfigure{\label{fig:c_delta} \includegraphics[scale=0.55]{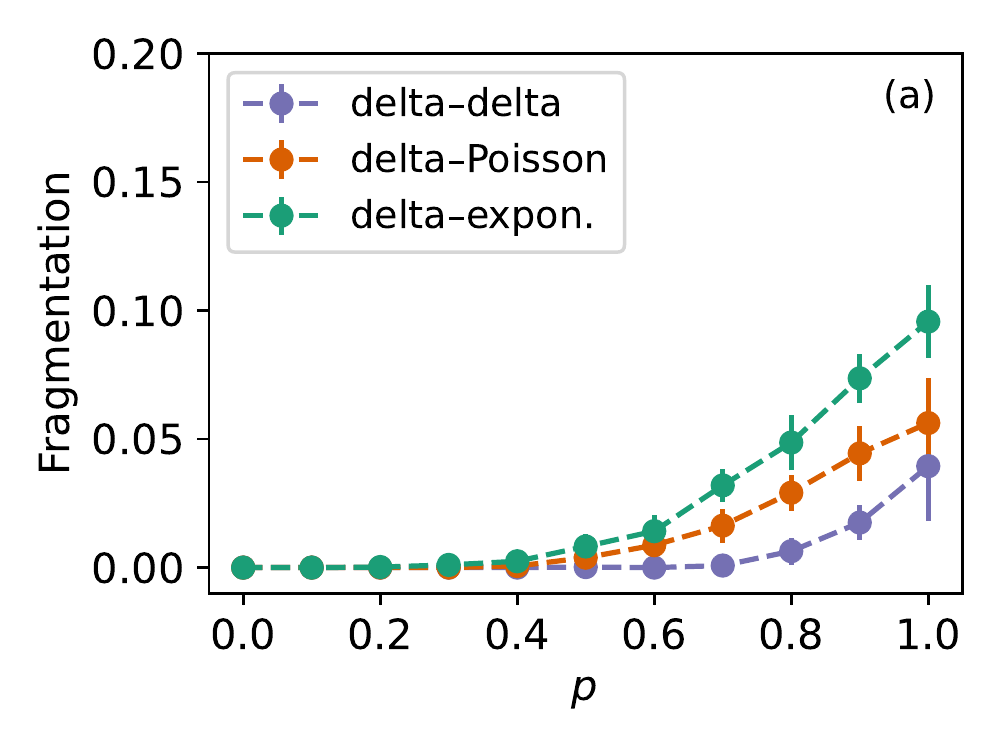}}
	\subfigure{\label{fig:c_poisson} \includegraphics[scale=0.55]{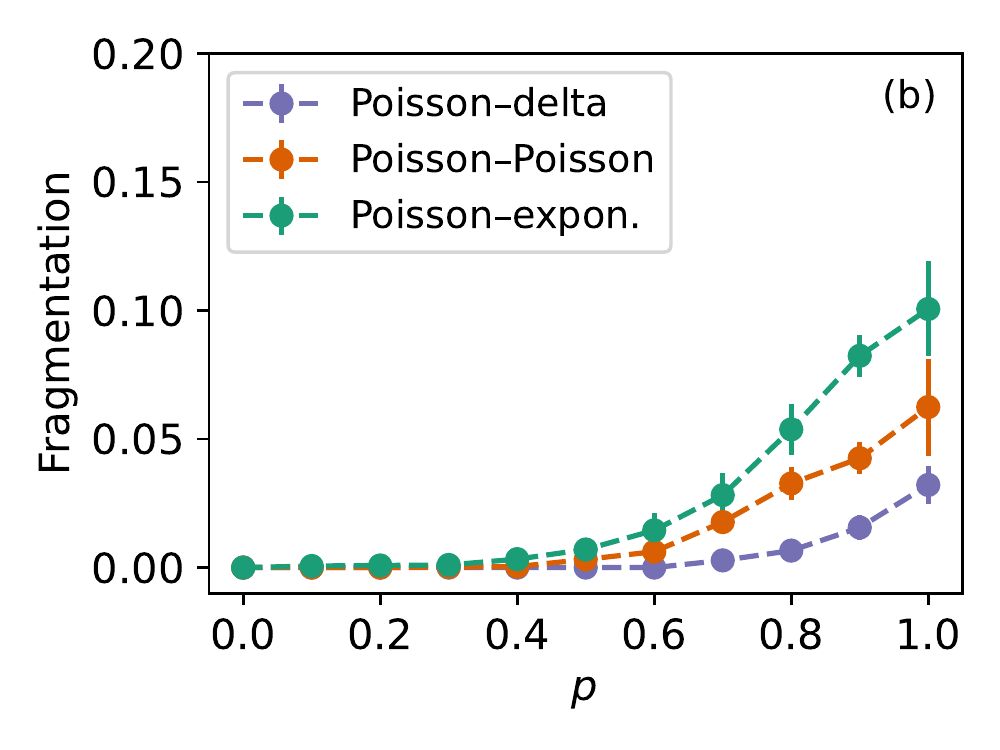}}
	\subfigure{\label{fig:c_exp} \includegraphics[scale=0.55]{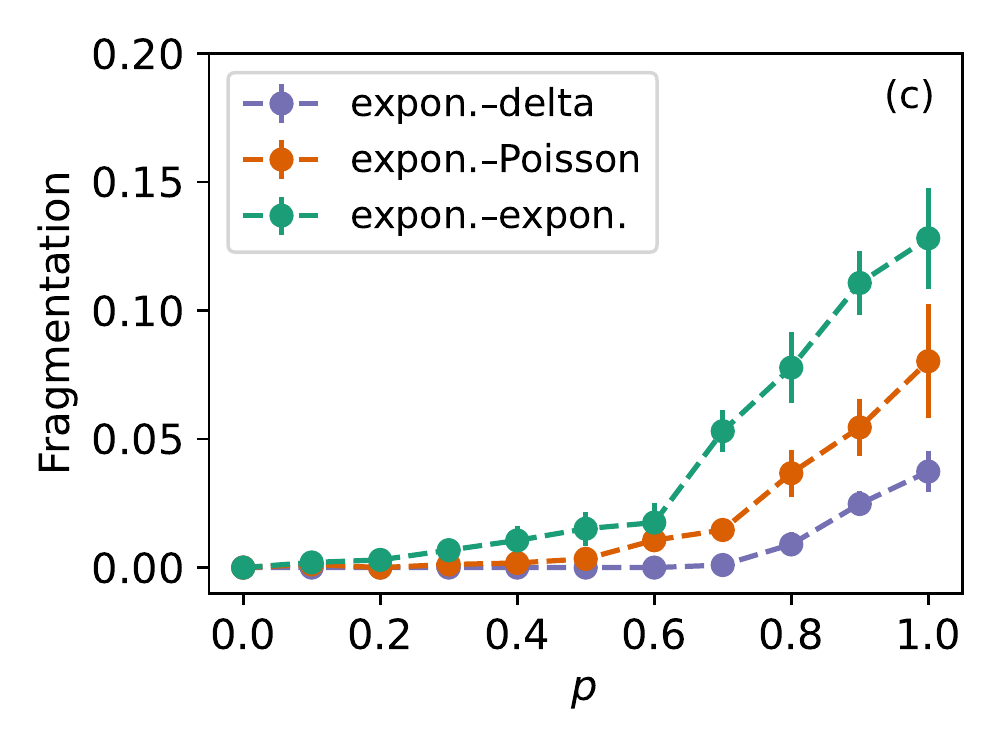}}
	\caption{Fragmentation as a function of $p$ for all $P_{e}(d)$-$P_{u}(k)$ combinations with (a) delta function, (b) Poisson, and (c) exponential hyperedge size distributions. The fragmentation starts with lower $p$ values when both $P_{e}(d)$ and $P_{u}(k)$ are heavy-tailed. Values are the mean over ten runs, and bars are the standard deviation.}
	\label{fig:fragmentation_vs_p}
\end{figure*}

The networks start to fragment after a certain value of $p$, which is higher for peaked degree and hyperedge size distributions and lower for broader distributions (Fig.~\ref{fig:fragmentation_vs_p}). Initially, the fragmentation is very small with a few and small disconnected components. However, it increases fast after $p=0.6$ in all cases, indicating some transition at that level of overlapping for the artificial higher-order networks with the density we use here.   

What causes fragmentation is that, as overlapping increases, many edges, that would otherwise connect larger sets of (unique) nodes, repeatedly connect nodes from a smaller set. In other words, overlaps reduce edge availability for bridging the network together. At first glance, this result might seem contradictory to that of the clustering coefficient: with more overlaps, we have concurrent higher cohesion and fragmentation. However, on closer inspection, it becomes evident that, with the same density, more cohesive networks are more segregated, with possible fragmentation. That is, tightly-knit node subsets emerge, developing into communities, with less frequent or absent connections with other subsets.

To support this idea, we can also investigate how network modularity behaves as a function of the tuning parameter $p$. For that, let us use the Louvain method to detect communities~\cite{blondel2008fast}, and network modularity~\cite{newman2004finding} to calculate how well defined communities are, which helps illustrate network segregation.

\begin{figure*}
	\centering
	\subfigure{\label{fig:mod_delta} \includegraphics[scale=0.55]{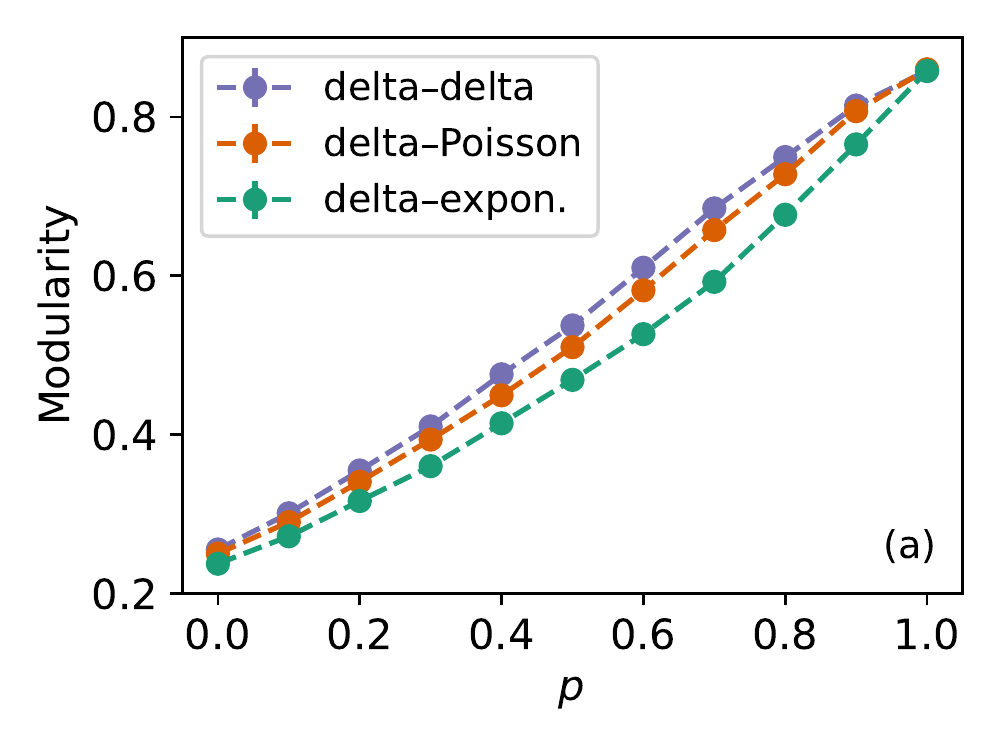}}
	\subfigure{\label{fig:mod_poisson} \includegraphics[scale=0.55]{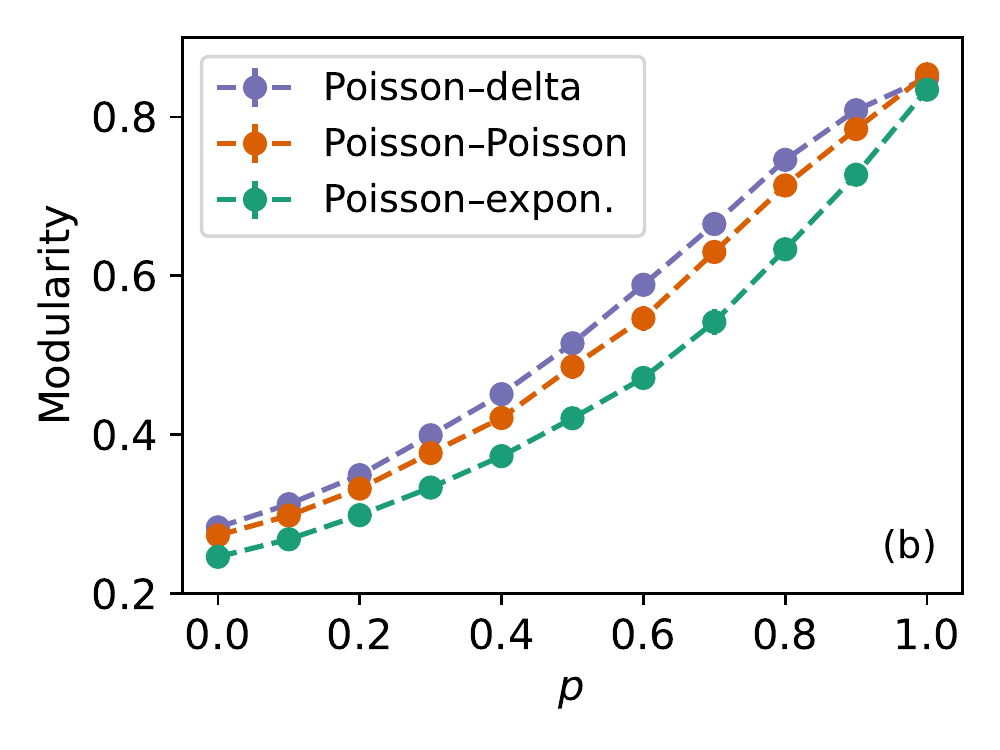}}
	\subfigure{\label{fig:mod_exp} \includegraphics[scale=0.55]{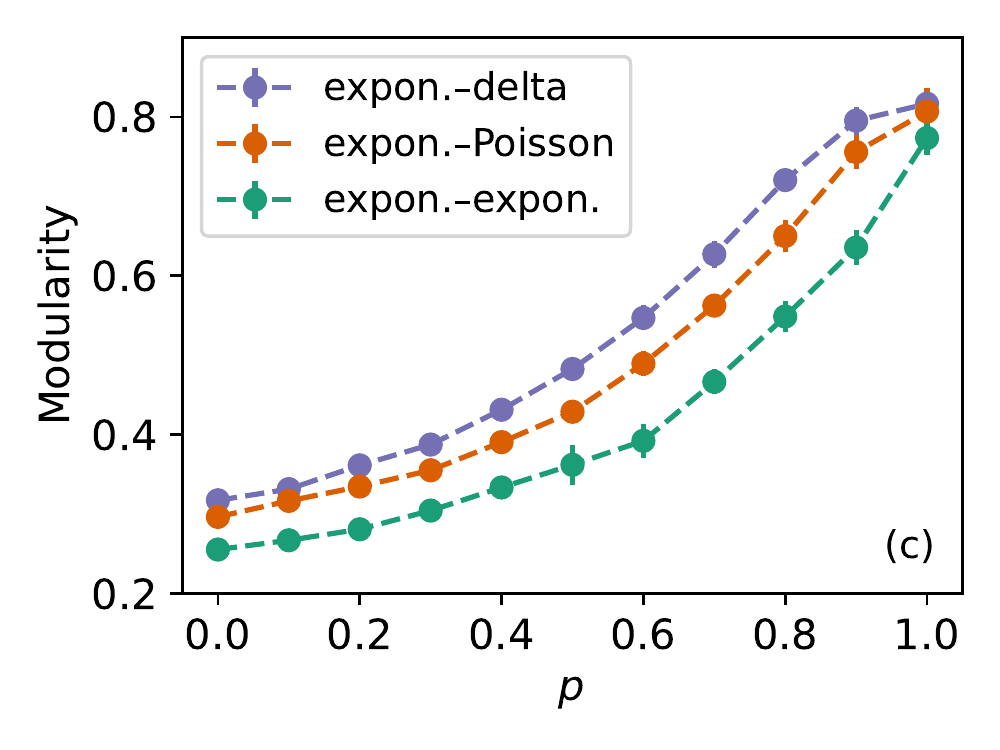}}
	\caption{Modularity as a function of $p$ for all $P_{e}(d)$-$P_{u}(k)$ combinations with (a) delta function, (b) Poisson, and (c) exponential hyperedge size distributions. The number of overlaps significantly increases modularity, segregating the network in well-defined communities, especially  when $P_{e}(d)$ is peaked. Values are the mean over ten runs. Standard deviation bars are unnoticeable.}
	\label{fig:modularity_vs_p}
\end{figure*}

As expected, modularity grows fast with $p$, more so with peaked degree and hyperedge size distributions, especially the latter (Fig.~\ref{fig:modularity_vs_p}). On the other had, broad hyperedge size distributions curb growing modularity because they create large subsets of connected nodes with individual large-sized hyperedges. For that exact reason, they result in networks with higher modularity for $p=0$. However, with raising overlapping levels, it takes longer for small-sized hyperedges connecting each of these nodes (in large subsets) with nodes from other subsets to start connecting nodes within the large subsets instead. 

To conclude, we have seen that the level of overlapping can drastically affect the structure of high-order networks in several ways. In particular, at high levels, it boosts segregation, with the emergence of well-defined community structures and leading to an eventual fragmentation of the network. While it seems counter-intuitive that overlaps increase both cohesion and segregation, the fact is that these two characteristics reinforce each other in networks with the same density. These findings are especially relevant to social systems, where understanding the mechanisms that result in more overlaps---and societal segregation---becomes critical.

%\section{Conclusions}

%\begin{acknowledgments}
%The authors would like to thank Chakresh Singh for helpful conversations. 
%\end{acknowledgments}

\bibliographystyle{unsrtnat}
\bibliography{fragmentation} 
% The references (bibliography) information are stored in the file named "Bibliography.bib"

%Significance statement
%By understanding the mechanisms through which transitivity and degree assortativity emerge

\end{document}